\documentclass{article}

\usepackage{PRIMEarxiv}

\usepackage{amssymb}

\usepackage{amsmath}

\usepackage{graphicx}
\usepackage{subcaption}
\PassOptionsToPackage{hyphens}{url}\usepackage{hyperref}

\usepackage{algorithm}
\usepackage{algpseudocode}
\usepackage[table]{xcolor}

\usepackage{booktabs}
\usepackage{multirow}
\usepackage{makecell}
\usepackage{tabularx}
\usepackage{array}

\newcolumntype{L}{>{\raggedright\arraybackslash}X}
\newcommand{\subrow}[1]{\hangindent=1em\hangafter=0 #1}

\title{Why Trust Your Agent? Empirical Security Gains from TRiSM-Guided Agentic Workflows in Healthcare
}

\author{
  Liam Kearns \\
  AuraQ \\
  Malvern \\
  Worcestershire \\
  UK \\
  \texttt{liamkearnsy@gmail.com} \\
}

\begin{document}
\maketitle

\begin{abstract}
Agent-based AI has enabled the automation of tasks by exposing application tools and resources to large language models (LLMs). However, to improve scope and accuracy, agents are often given access rights that exceed those of ordinary users, introducing significant security risks and potential regulatory breaches. This paper applies the AI Trust, Risk, and Security Management (TRiSM) framework to a medical report-generation application to demonstrate how an insecure baseline workflow can be transformed into a security-conscious agentic workflow. Both designs were evaluated with five LLMs (Claude Haiku 4.5, GPT-4.1-nano, GPT-4.1-mini, GPT-5.4-mini and Gemini 2.5 Flash) across 800 report generations and 500 attack trials covering RAG poisoning, data-field injection and client-side network injection. The redesign reduced attack success from 31\% to 10\% for RAG poisoning and from 42\% to 25\% for data-field injection (McNemar p $< 0.001$ for both), removed network injection by construction, and raised report accuracy from 72.5\% to 86.5\%. In an exploratory post-hoc grouping by attack goal, success against attacks aimed outside the report task, such as data disclosure and tool abuse, fell from 21.7\% to 2.5\% and no patient record was disclosed under the redesign. Attacks altering the reported Covid-19 diagnosis remained effective, accounting for 32 of the 35 successful attacks against the redesign. Least-privilege architecture limits what an injection can reach, but not whether a model follows injected content within its task, so clinical outputs need verification independent of the LLM. No model resisted all three attack vectors, so model selection cannot substitute for secure workflow design.

\end{abstract}




\section{Introduction}
\label{sec:intro}

Agentic applications have become a popular way to infuse AI capabilities into projects. Agents, often powered by LLMs, work more effectively when provided with sufficient data sources and application tools. While this allows agentic AI to be used in high-value settings such as medical diagnosis, it raises patient safety concerns, because these automated decisions often lack accountability and transparency \cite{Sarfraz2025}. Broad access also exposes sensitive data, both to the providers hosting the models and to users who can manipulate the requests sent to those models and so view the sensitive information they contain.

Recent advances have expanded the capabilities of agentic workflows. Model Context Protocol (MCP) servers provide structured protocols in which agents can invoke tools and execute actions, while Retrieval-Augmented Generation (RAG) with vector knowledge bases gives agents access to context-specific information at inference time. However, these components create a greater attack surface for agentic workflows. MCP tool logic can be opaque to the integrating application, making it difficult to map out security risks. Similarly, knowledge bases are not guaranteed to comply with the same access controls as application databases, increasing the risk of sensitive information being compromised.

Integrating AI into applications should follow a framework that manages its risks. Without one, insecure practices such as inexperienced vibe coding, where AI-generated code is integrated with minimal review, can produce code that works but is open to injection and insecure design \cite{Songwen2026}. To address this, frameworks such as AI Trust, Risk, and Security Management (TRiSM) can structure model governance and trustworthiness.

This paper investigates the vulnerabilities introduced by improper agent integration and how TRiSM-guided design can reduce them. Security challenges from the complexity of agent execution logic have been highlighted by previous research \cite{Deng2025}, but mitigations are rarely evaluated empirically in working applications.  The motivation for this work came from an existing application that called LLMs through basic REST API logic with minimal guardrails, putting both the credibility of AI outputs and the security of the application at risk.

\begin{figure} [htbp]
    \centering
    \includegraphics[height=2.7in]{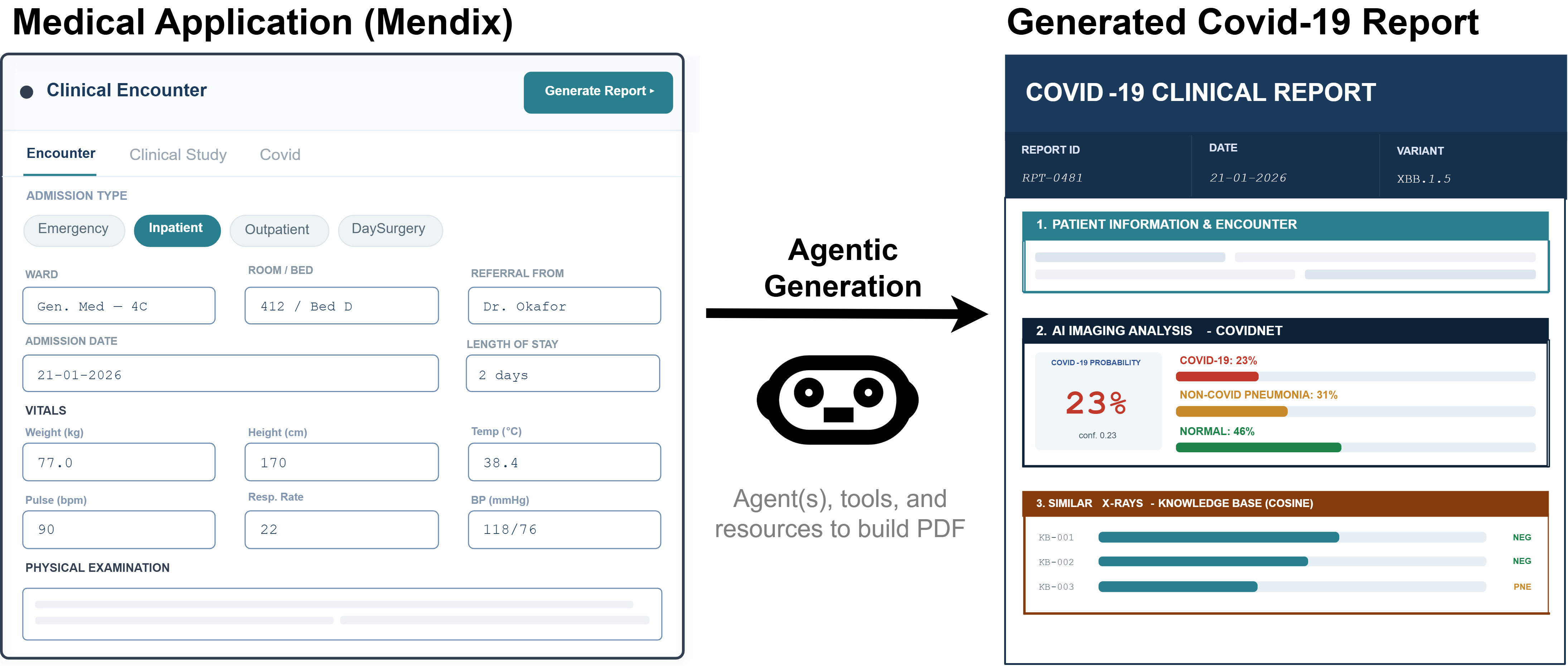}
    \caption{Medical application using AI agents to generate medical reports.}
    \label{fig:design}
\end{figure}

The scope of this paper is focused on using AI agents to generate medical reports in a medical application built with Mendix, a low-code application development platform. Figure~\ref{fig:design} illustrates clinical encounter information from the medical application using agents, tools, and resources to generate medical reports for regular and Covid-19 encounters. This application uses LLM generative capabilities to also generate recommended treatments if not provided by the attending physician. This scope provides a clear scenario to implement a secure agentic workflow to protect the application and patient data alongside the accuracy of generated reports.

The structure of the paper is as follows. Section \ref{sec:related} investigates significant attack vectors that AI-powered applications need to defend against. Section \ref{sec:threat} describes the threat model for the medical application and maps its attack vectors to the OWASP Top 10 lists for LLM and agentic applications, laying the foundation for Section \ref{sec:method} to outline the insecure baseline workflow and TRiSM-guided workflow as well as evaluation metrics. Next, Section \ref{sec:results} discusses the vulnerabilities and performance of both workflows across various LLM models. Subsequently, Section \ref{sec:reco} outlines the final recommended agentic workflow which is influenced by TRiSM principles.

\subsection{Contributions}
In this paper, vulnerabilities in AI-powered applications are identified and controlled through risk management. The paper makes the following contributions:

\begin{itemize}
    \item Applying TRiSM principles to a working medical agentic application, mapping each principle to a concrete architectural control.
    \item Demonstrating that a TRiSM-guided redesign reduces attack success while improving report accuracy. The gains are attributed to the redesign as a whole rather than to any single control.
    \item Showing which attacks the redesign does and does not stop. Attacks aimed outside the agent’s task are nearly eliminated, while attacks that alter the clinical conclusion remain effective.
    \item Providing per-model attack-success rates, reported as indicative, for agents granted access beyond the underlying user’s permissions.

\end{itemize}

\section{Related works}
\label{sec:related}
AI capabilities can be implemented in various ways within applications, each introducing its own risks. This section reviews the vulnerabilities most relevant to LLM-integrated medical applications and shows that existing mitigations address each vulnerability in isolation, but rarely in a combined architectural response for a medical application. Additionally, the AI Trust, Risk, and Security Management (TRiSM) framework is introduced that aims to mitigate these risks.

\subsection{Injection attacks}
Prompts sent to LLMs can be manipulated without attackers having access to model gradients or weights. These prompt injection attacks can be used for prompt leaking or code generation with a high success rate in LLM-integrated applications \cite{Gelei2025}. Even when mitigated through instructing the output to be a specific phrase or format (known-answer detection), LLMs used in decision making can still be heavily influenced by prompt injection \cite{Jiawen2024}. This highlights how injection attacks can impact LLM evaluations and bypass detection techniques in the output.

Injection attacks can be mitigated by enforcing input and output formats and time constraints on outputs \cite{Gelei2025}. However, indirect injection attacks can occur through attackers injecting prompts into data which will be used to build prompts \cite{Greshake2023}. Injection opportunities increase significantly when using RAG, where an attacker can subtly change semantics to introduce bias into retrieved information \cite{Gaurav2025}. With data and instructions both being used in prompts, injection attacks can be the consequence of traditional data integrity attacks such as data tampering and SQL injections. 

For medical applications, injection attacks can pose danger to patient confidentiality and physical wellbeing. RAG is often used in medical scenarios to provide a ground truth, but clinically phrased injections can influence treatment decisions or fabricate evidence \cite{Junhyeok2026}. This shows the need for safety controls at the orchestration layer of AI-powered workflows rather than expecting models to identify suspected injected data.

\subsection{Inference attacks}
Models can leak the data they were trained or fine-tuned on. Fine-tuning samples contributed by users can be inferred by other users \cite{Kandpal2023}, training data can be reconstructed from gradients \cite{Ligeng2019}, hidden prompts can be recovered from outputs \cite{Morris2023}, and membership of a dataset can be tested without access to weights \cite{Pratyush2024}. To reduce the risk of model inference, noise can be added while training the model. However, this reduces the accuracy of the model \cite{Ligeng2019}. This highlights the risk of reducing model accuracy to protect against data leaks.

Accuracy and confidentiality are essential for AI-powered medical applications. While synthetic health data aims to protect patient privacy, it is susceptible to inference attacks where legitimate patient data has been used to generate synthetic records \cite{Ziqi2022}. Medical images are also vulnerable to being reconstructed by attackers through inference. While model perturbation can reduce successful attacks, it can significantly reduce accuracy if too much noise is added to the model \cite{Maoqiang2020}. These attacks are outside the threat model of this paper (Section \ref{sec:threat}), but they motivate its emphasis on data minimisation, since patient data that never reaches a model cannot be inferred from it.

\subsection{Memory and knowledge-base poisoning}

Accepting unstructured inputs allows LLMs to be accessible to non-technical users but also makes them poor at separating chat history from new input. This allows tampering with chat history which is difficult to detect \cite{Cheng2024}. Tampering can be reduced through application-level controls to prevent users from directly interacting with knowledge bases that provide inputs to LLM agents. Furthermore, dynamic permissions can be implemented so that only knowledge bases relevant to the current task are accessible \cite{Tianneng2025}. However, if knowledge bases are produced from external or untrusted sources, attackers can inject malicious instances to poison the database \cite{Zhaorun2024}. This highlights that inputs from untrusted sources can lead to poisoning agent context if appropriate access controls are ignored.

The risk compounds when knowledge bases also serve as semantic memory for agents. In multi-agent systems, poisoned memory can propagate between agents \cite{Torra2026}, creating an attack surface that is not present in single-agent systems. Multi-defence frameworks exist to reduce injection risks through content filtering and response verification. However, attacks such as instruction overrides still have an attack success rate of over 10\% with this multi-defence framework \cite{Badrinath2025}. This indicates that attacks through untrusted knowledge bases can still evade layered defences, requiring defence-in-depth at the framework level.

\subsection{Tool misuse}

Giving agents direct access to application databases can improve response accuracy, but expands the attack surface significantly. Unrestricted access to a database creates a new entry point via agents, whereby any vulnerability in agent logic can become a path to unauthorised data exposure \cite{Raihan2024}. Because of this new entry point, agents should be viewed as a new user role within applications; agents access tools and data to complete tasks, therefore requiring the same access controls and logging applied to human users to prevent misuse of application functionality or data.

Providing agents with tools allows them to complete external actions instead of being confined to internal environments. However, attackers can store malicious data on these external systems to compromise the internal application through the agent \cite{Ang2025}. With these attacks being possible through links that aim to download malware or direct to fake payment forms, attackers do not need to understand how the underlying AI model works and can instead use the same attack mechanisms designed for human users. Model Context Protocol (MCP) extends this exposure further by providing an open standard for agents to consume external tools, resources, and prompts \cite{Xinyi2025}. With the protocol simplifying connections to multiple external sources, risk is amplified through an increased attack surface which can be exploited through untrusted MCP servers \cite{Herman2025}. By using external black box MCP servers, systems can be exposed to malicious code operations which agents can execute autonomously. This demonstrates the need for tool restrictions through access controls to reduce application exposure.

\subsection{Hallucination and trust}

A challenge with agentic AI integration is the trust in outputs; LLMs are prone to hallucinations, requiring caution when reviewing their outputs. Knowledge bases can provide ground truth through RAG, but introduce new concerns for data governance: which agent is accessing what data, and with whom the outputs are shared \cite{Sarfraz2025}. Building secure data governance can be challenging with published agentic systems often lacking transparent security metrics. Some of these agentic systems use agent builders that delegate safety responsibilities to users, raising concerns that safety-critical behaviours could be undermined \cite{Leon2026}. Without an explicit framework for cyber security resilience, integrating LLMs into applications increases risk through novel threats. If the AI makes a harmful or incorrect decision, public trust in the system can be broken \cite{Vatamanu2025}.

Vulnerabilities caused by agents are often due to the trust provided to these agents; the more sections of an application an agent is given unrestricted access to, the greater the attack surface for agent misuse. Moreover, unintentional access to agents can be given through insecure code. This can be exacerbated through vibe coding the agent logic where such code has demonstrated high instances of exploitable vulnerabilities \cite{Songwen2026}. Such logic can increase maintenance and unpredictability of codebases \cite{Marko2025}. Therefore, guardrails should be implemented to prevent excessive autonomy of agents which risk hallucination and misaligned goals \cite{Shaina2025}. This highlights the need to control agent trust to reduce the risk of attack on AI-augmented applications.

Despite the risks of data exposure and output hallucinations, excessive trust is often put into AI-powered workflows. Data minimisation can reduce sensitive data exposure to LLMs while maintaining model performance in many cases, but users and systems often overshare irrelevant data \cite{Jijie2025}. This highlights the need to implement agents with defined boundaries, where permissions, prompt construction, and tool exposure are governed for safe deployment. The TRiSM framework provides the structure for that secure workflow.

\subsection{Architectural defences for agents}
Defences built into the system around a model can be more dependable than relying on the model to resist injection. Design patterns have been proposed that deliberately constrain what an agent can do once it has read untrusted input, so that the input cannot trigger harmful actions or produce harmful outputs \cite{Beurer2025}. CaMeL separates the control flow derived from the trusted user query from the data the agent retrieves, so that retrieved content cannot change the program’s flow, and tracks capabilities on each value to block unauthorised data flows \cite{Debenedetti2026}. On the AgentDojo benchmark \cite{agentdojo2024}, it solved most tasks with provable security, at some cost in utility. These defences mainly target injected content that triggers actions or leaks data. They have rarely been evaluated in clinical applications, where the most consequential injected content may change what a report says rather than what the agent does.

\subsection{TRiSM framework}
The TRiSM framework was created by Gartner to address governance and risk in AI systems. TRiSM's emphasis on transparency and monitoring is intended to help systems adapt to new risks \cite{Adib2024}. Standards such as ISO/IEC 42001 are suited to the organisational level, but without actionable controls, creating technical implementations can be a challenge \cite{Pamela2025}. With TRiSM being a flexible framework, technical implementations can be created at the agent level, which is beneficial for agentic workflows with non-deterministic outputs.

To protect an organisation's increased attack surface from generative AI, TRiSM focuses on five principles: Governance, Trustworthiness, Fairness, Reliability, and Data protection \cite{Gomstyn2025}. These principles aim to provide transparency and explainability within AI systems to mitigate against adversarial inputs and unsafe outputs. Models used for agents evolve continually, producing expanding capabilities and changing governance considerations; TRiSM assists with managing AI risk in this dynamically governed environment \cite{Ray2026}. This makes the framework suitable for workflows which require constant updates.

Existing research has explored TRiSM conceptually to address risks from LLM models and governance. The framework can control attack surfaces from agent communication and collaboration with agentic AI systems \cite{Shaina2025}. While AI TRiSM has been noted as beneficial in healthcare \cite{Adib2024,Gomstyn2025} and can protect sensitive health information \cite{Adib2024}, empirical research applying TRiSM to implement architectural controls within working medical applications remains limited. This paper contributes an empirical application of TRiSM to a working medical agentic workflow, mapping each TRiSM principle to the workflow and quantifying the resulting security and performance trade-offs.

\section{Threat model}
\label{sec:threat}
This section defines what the workflows protect, who the attackers are and what they can do. The baseline application already enforced access controls in its database and user interface, but these were designed for human users; an agent that assembles prompts from several data sources creates paths around them. The threat model therefore starts from the assets an agent can reach and asks which of them an attacker could read or alter through the agent.

\subsection{Assets}
The workflows have four assets which are identified as potential targets of an attack: patient records, including personally identifiable information; the integrity of the generated report, in particular the Covid-19 determination and probability; hidden context such as system prompts and tool schemas; and the knowledge base of reference X-rays. Report integrity is the asset most specific to clinical use. An altered report looks like a normal one, so the error may go unnoticed before it influences care.

\subsection{Adversaries}
In the workflows, three adversaries are considered and tested against. Table \ref{tab:adver} outlines these adversaries and which attack vectors they have within the medical application. A fourth adversary, the operator of a malicious or compromised MCP server, was not tested. Without access control on tools imported from MCP servers, every tool a server offers can be passed to the agent, so an attacker who controls the server can supply tools that the agent executes autonomously \cite{Herman2025}. The baseline gives its single agent every available tool. The TRiSM-guided workflow limits each agent to the tools listed in its policy JSON, so tools a compromised server adds are blocked by default. The policy controls which tools an agent may call, however, not what those tools do, so a server that changes the description or behaviour of an allowed tool would not be detected. Defending against that requires pinning or verifying tool definitions, which this study did not implement.

In all cases the adversary has no access to model weights, gradients, provider infrastructure or the application server itself. Each attack has one of four goals: altering the clinical conclusion of the report, disclosing data or hidden context, causing unintended tool use, or replacing the report output.

\begin{table}[htbp]
\centering
\caption{Adversaries considered in the threat model.}
\label{tab:adver}
\begin{tabularx}{\textwidth}{>{\raggedright\arraybackslash}p{3.6cm}
                              >{\raggedright\arraybackslash}p{3.0cm}
                              >{\raggedright\arraybackslash}X}
\toprule
\textbf{Adversary} & \textbf{Vector} & \textbf{Capability assumed} \\
\midrule
Authenticated application user &
Network injection &
Modifies the client-side request state sent to the server, for example with an intercepting proxy such as Burp Suite \\
\addlinespace
Writer of clinical free-text fields &
Data-field injection &
Writes text into encounter fields that are later included in prompts, for example through a staff account, an upstream system, or a compromised account \\
\addlinespace
Writer of knowledge-base content &
RAG poisoning &
Writes text into a record returned by similarity search, for example through a compromised ingestion pipeline or insider access \\
\bottomrule
\end{tabularx}
\end{table}

Attack vectors and goals identified in the medical application are listed in Table \ref{tab:att_vec} and mapped to the OWASP Top 10 for LLM Applications \cite{OWASP2026}. Inference attacks that require model internals, training-data extraction, compromise of the LLM provider, and denial of service are out of scope, although Section \ref{sec:related} discusses several of them. Because the application’s agents call tools, each row is also mapped to the OWASP Top 10 for Agentic Applications \cite{OWASP2025}. RAG poisoning is mapped to LLM01 rather than LLM09, since the payloads are instructions carried in retrieved content and do not depend on the embedding layer.

\begin{table}[htbp]
\centering
\caption{Attack vectors and goals mapped to the OWASP Top 10 for LLM Applications and for Agentic Applications.}
\label{tab:att_vec}
\begin{tabularx}{\textwidth}{>{\raggedright\arraybackslash}X
                              >{\raggedright\arraybackslash}p{4.2cm}
                              >{\raggedright\arraybackslash}p{4.2cm}
                              c}
\toprule
\textbf{Attack vector or goal} & \textbf{OWASP LLM Top 10} & \textbf{OWASP Agentic Top 10} & \textbf{Tested} \\
\midrule
\multicolumn{4}{l}{\textit{Attack vectors}} \\
\addlinespace[2pt]
Network injection    & LLM01 Prompt Injection & ASI01 Agent Goal Hijack & Yes \\
Data-field injection & LLM01 Prompt Injection & ASI01 Agent Goal Hijack & Yes \\
RAG poisoning        & LLM01 Prompt Injection & ASI06 Memory and Context Poisoning & Yes \\
Malicious MCP server & LLM04 Supply Chain & ASI04 Agentic Supply Chain Vulnerabilities & No \\
\midrule
\multicolumn{4}{l}{\textit{Attack goals}} \\
\addlinespace[2pt]
Diagnostic manipulation & LLM07 Misinformation (induced via LLM01) & ASI09 Human-Agent Trust Exploitation & Yes \\
Disclosure of data or hidden context & & & \\
\subrow{Patient records} & LLM02 Sensitive Information Disclosure & ASI03 Identity and Privilege Abuse & Yes \\
\subrow{System prompt or tool schema} & LLM08 Hidden Context Exposure & --- & Yes \\
\subrow{Through embedded URLs (e.g.\ tracking images)} & LLM10 Improper Output Handling & --- & Yes \\

Unintended tool use & LLM03 Excessive Agency & ASI02 Tool Misuse and Exploitation & Yes \\

Output hijacking or substitution & LLM01 Prompt Injection & ASI01 Agent Goal Hijack & Yes \\
\bottomrule
\end{tabularx}
\end{table}

\section{Methodology}
\label{sec:method}

Although both AI agents and agentic AI typically use LLMs, their architecture can differ substantially. AI agents are suitable for isolated single-step tasks, whereas agentic AI can divide a task into subtasks for collaboration between multiple specialised agents \cite{Ranjan2026}. The agentic workflow provides a suitable foundation to apply TRiSM principles; data partitioning improves data minimisation, single-call agents reduce the memory-poisoning surface, and per-agent tool scoping enforces least privilege. This paper compares the baseline and TRiSM-guided workflows, both applied to the same medical application, highlighting the importance of designing a secure framework based on the application's goals and risk profile.

\begin{figure} [htbp]
    \centering
    \includegraphics[height=2in]{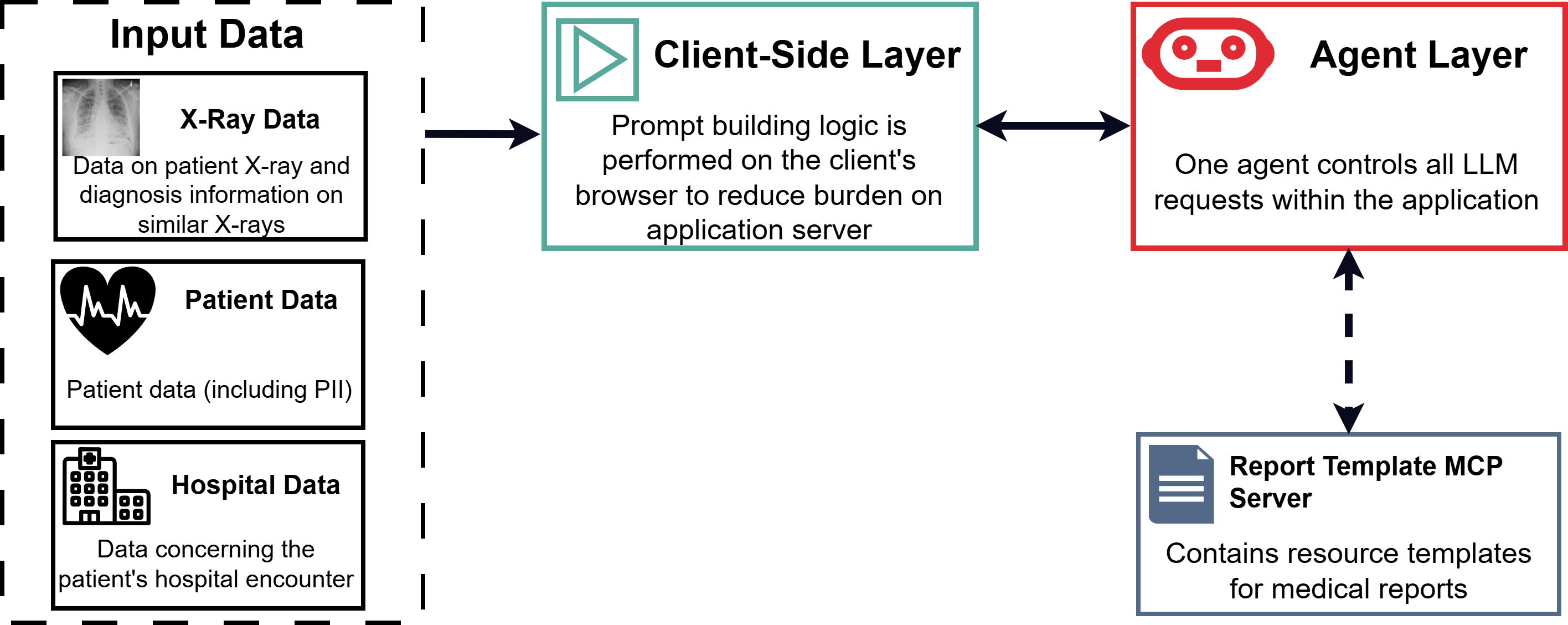}
    \caption{Baseline workflow for medical report generation application.}
    \label{fig:agent}
\end{figure}

\begin{figure} [htbp]
    \centering
    \includegraphics[height=2in]{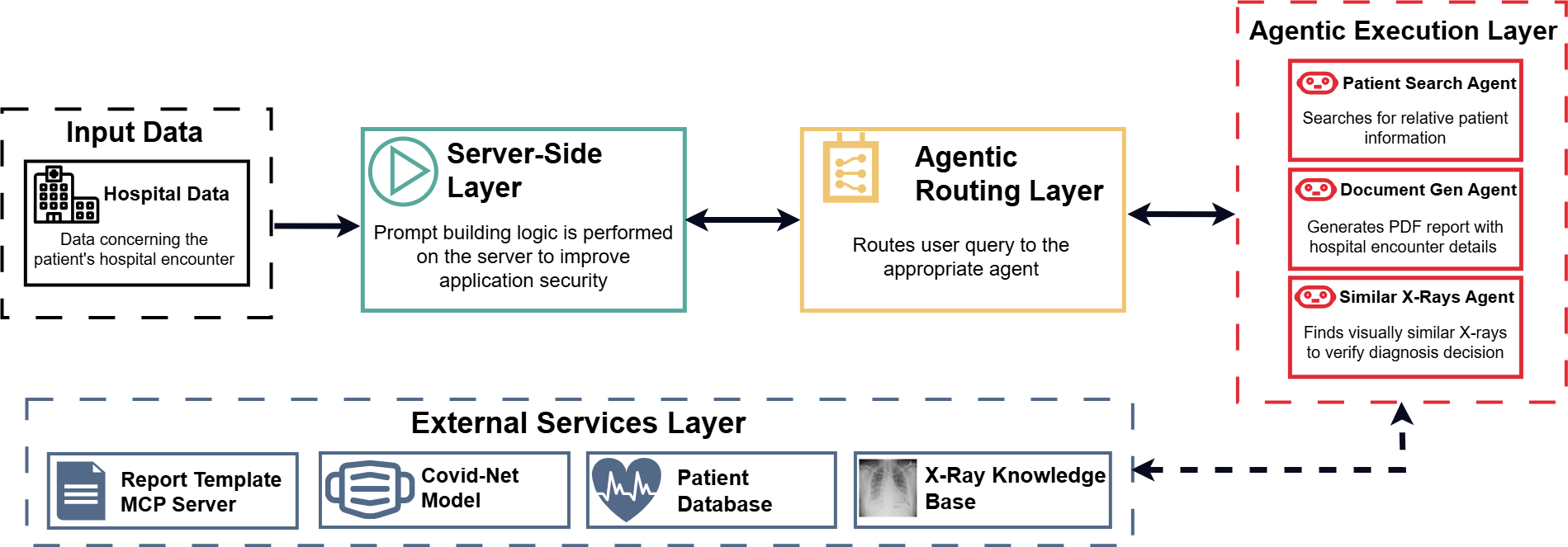}
    \caption{TRiSM-guided workflow for medical report generation application.}
    \label{fig:agentic}
\end{figure}

The baseline in Figure \ref{fig:agent} reproduces the application’s original design. Although the database and user interface enforce access controls, the agent’s prompts are built from client-side copies of the data, so those controls apply neither to what the agent receives nor to what a user can change. Figure \ref{fig:agentic} shows the TRiSM-guided workflow, which divides the task between specialised agents.

\subsection{Agents and controls used}
The proposed TRiSM-guided workflow utilises multiple specialised agents, each with access to a set of tools and resources. This contrasts with the baseline workflow, whereby a single agent has access to all tools and resources.

The Router agent receives the report request and determines which agents and tools are required for the encounter type. It does not access patient data.

The Patient Search agent retrieves patient records relevant to the hospital encounter. It outputs a JSON of information required by the downstream agents, ensuring sensitive information has a reduced exposure to only agents that require it. 

The Document Generation agent produces the medical report in PDF format. It receives hospital encounter and patient information in addition to Covid-19 information depending on the encounter type. It accesses tools and resources from an external MCP server to apply the appropriate report template for the type of hospital encounter.

The Similar X-ray agent is used exclusively for Covid-19 hospital encounters. It queries the X-ray knowledge base using the patient's X-ray image to find diagnosis statuses for similar X-rays. Furthermore, it runs Covid-Net \cite{WangLinda2020}, a discriminative convolutional network for detecting Covid-19 in chest X-rays. By obtaining a predicted diagnosis and analysing diagnoses from similar X-rays, the reliance on generative AI alone for clinical judgement is reduced.

Three further controls surround the agents, as outlined in Figure \ref{fig:flow-secure}. Embedding analysis is a hybrid detector applied to the hospital encounter’s report request, including the encounter’s text fields, before inference. The text is embedded with OpenAI’s text-embedding-3-small model and scored in three ways. A semantic score is its highest cosine similarity to a reference set of 55 injection phrases in seven families: instruction override, role hijacking, delimiter confusion, data exfiltration, content manipulation, goal hijacking, and indirect injection. An anomaly score is its distance from the centroid of benign examples, relative to the typical distance of those examples from the centroid, scaled so that three times the typical distance scores 1. A lexical score adds a fixed increment for each of five categories of injection keywords found in the text, up to 0.6 in total. The final risk is 0.65 times the semantic score plus 0.35 times the anomaly score plus the lexical score, capped at 1. Requests with a risk of 0.70 or above are rejected and logged; requests scoring between 0.40 and 0.70 are logged as suspicious and allowed. No legitimate report request in the accuracy evaluation was rejected by the detector.

A policy JSON lists the tools each agent may call; calls not listed are blocked by default and logged. Response analysis checks that no unauthorised tools were called before an agent’s output is passed on or returned.

\subsection{Framework used}
\label{sub:framework}
AI TRiSM is a framework for Trust, Risk, and Security Management in AI systems. With this framework designed to handle risks and governance, applying the framework to applications can help with regulatory compliance \cite{Adib2024}. Furthermore, the framework can help monitor agent performance and reduce exposing sensitive medical data \cite{Ray2026}. As this paper focuses on a medical application, regulatory compliance is crucial for patient safety. This framework guided the security-conscious workflow within this paper's medical application. Table \ref{tab:trism} maps each TRiSM principle to the control implemented in the TRiSM-guided workflow. The evaluated workflow contains all of these controls together; the last column states whether a control’s effect was isolated through its own logs, measured only as part of the redesign, or not measured.

\begin{table}[htbp]
\centering
\caption{TRiSM principles implemented into the medical application.}
\label{tab:trism}
\small
\renewcommand{\arraystretch}{1.15}
\begin{tabularx}{\textwidth}{@{}>{\raggedright\arraybackslash}p{2.15cm}
  >{\raggedright\arraybackslash\hsize=1.05\hsize}X
  >{\raggedright\arraybackslash\hsize=0.85\hsize}X
  >{\raggedright\arraybackslash\hsize=1.10\hsize}X@{}}
\toprule
\textbf{Principle} & \textbf{Control} & \textbf{Threat addressed} & \textbf{Evaluated} \\
\midrule
Data protection &
Server-side prompt construction; data partitioning between agents; per-agent tool scoping enforced by a policy JSON, with default block and fallback &
Network injection; data disclosure; excessive agency &
Policy JSON: attributable through its logs (6 of 43 fixed trials; the model attempted each blocked call).\newline
Prompt construction and partitioning: as part of the redesign \\
\addlinespace[4pt]
Fairness &
Data minimisation of prompt content; embedding analysis before inference &
Injected content biasing outputs &
Embedding analysis: attributable through its logs, upper bound (at most 6 of 43 fixed trials).\newline
Output fairness: not measured \\
\addlinespace[4pt]
Reliability &
Prompt state held server-side, with no client write access; response analysis with content filtering &
Tampering with request state; unsafe outputs &
As part of the redesign; response analysis did not log its decisions \\
\addlinespace[4pt]
Governance &
Per-agent versioning; single-call agents &
Cross-request leakage through shared conversation; memory poisoning; uncontrolled agent changes &
Single-call agents: as part of the redesign (cross-request disclosure).\newline
Memory poisoning and versioning: not measured \\
\addlinespace[4pt]
Trustworthiness &
Logging of agent and service calls; non-generative evidence from Covid-Net and similar-X-ray retrieval &
Hallucinated or unverifiable reasoning &
Not measured \\
\bottomrule
\end{tabularx}
\end{table}

The medical application was developed using Mendix Studio Pro version 10.24, and agents were implemented using Mendix Agent Commons, which supports version control and integration of LLM models via API calls. The AI Agent Index highlights that most agents depend on GPT, Claude, or Gemini, resulting in dependency on a small number of providers and a single point of failure  \cite{Leon2026}. Using Agent Commons mitigates this risk by allowing alternative model providers to be used without architectural changes. All model calls were made via direct provider APIs (Anthropic, OpenAI, and Google) using each provider's default decoding parameters; experiments were conducted in May-June 2026 using the following model versions: claude-haiku-4-5-20251001, gpt-4.1-nano-2025-04-14, gpt-4.1-mini-2025-04-14, gpt-5.4-mini-2026-03-17, and gemini-2.5-flash.
\footnote{%
gemini-2.5-flash is Google's stable identifier for the Gemini 2.5
Flash release of 17 June 2025, which was unchanged from the gemini-2.5-flash-preview-05-20 snapshot}

\subsection{Report generation}
All patient and encounter records were synthetic, created for this study and contain no real patient data. The X-ray images associated with encounters are from the COVIDx CXR-3 dataset \cite{Pavlova2022}, as are the reference X-rays in the knowledge base. Encounter images were excluded from the knowledge base. Two report types were generated: regular hospital encounters and Covid-19 encounters, which require additional tools and a more complex template. This provides comparison on how the workflows manage tasks which require different tools and resources. Five LLMs were tested across both workflows: Claude Haiku 4.5, GPT-4.1-nano, GPT-4.1-mini, GPT-5.4-mini, and Gemini 2.5 Flash.

Each report type was generated 40 times per model and workflow, giving 800 generations. A generation was scored accurate if it met all four criteria: the workflow called the correct tools, completed all mandatory fields, included no hallucinated report content, and represented the results of Covid-19 tools correctly where applicable. Latency was measured from the report request to the finished PDF, and cost was computed from token counts at each provider’s published rates at the time of the study. Reasoning tokens that Gemini 2.5 Flash reports separately from its response were billed at the output rate. Runs that ended in an error before producing a report were counted as inaccurate and excluded from token, latency and cost figures. There were 31 such runs, 24 under the baseline and 7 under the TRiSM-guided workflow.

\subsection{Attack set and scoring}
The attack set was built for the medical application rather than taken from a general-purpose benchmark such as AgentDojo \cite{agentdojo2024}, whose task suites are not clinical and may not capture outputs that are policy-safe but clinically dangerous \cite{Junhyeok2026}. The attack set contains 20 scenarios for each of the three vectors:
\begin{itemize}
  \item\textbf{Network injections} were performed as an authenticated user, using Burp Suite as an intercepting proxy to modify client-side state in requests before they reached the application server.
  \item\textbf{Data-field injections} were inserted into free-text encounter fields.
  \item\textbf{RAG poisoning payloads} were added to the metadata of X-ray records returned by similarity search.
\end{itemize}

Each scenario was run once per model and workflow. Network injection applies only to the baseline, because client-supplied fields never enter server-side prompt construction. The baseline therefore received 60 attacks per model and the TRiSM-guided workflow 40, giving 500 in total.

 An attack was recorded as successful if a model's output deviated from the attack-free ground-truth output in the direction of the intended injection. This includes attacks such as altered or suppressed Covid-19 determination, inclusion of data or fields the injection requested, execution of an injected tool-calling instruction, and substitution of report content with attacker-specified text. To examine which attacks the redesign stops, each RAG and data-field scenario was also assigned, from its payload text, to one of the four attack goals in Section \ref{sec:threat}. These were diagnostic manipulation (16 scenarios), disclosure of data or hidden context (14), unintended tool use (5) and output hijacking or substitution (5). Scenarios with more than one aim were assigned to their primary goal.

\subsection{Statistical analysis}
Attack success rates are pooled over the five models (100 trials per vector and workflow) and reported with 95\% Wilson score intervals. Because each RAG and data-field scenario ran under both workflows, workflow differences are tested with McNemar’s exact test on discordant pairs. The same 20 scenarios were replayed across models, so pooled trials are not fully independent. As a robustness check, each scenario’s number of successful models (0 to 5) is compared between workflows with a two-sided sign test. Accuracy is reported with 95\% Wilson score intervals. Per-model results are indicative only: with 20 attacks per cell, a 20\% success rate has a 95\% interval of roughly 8\% to 42\%.

\subsection{Scope of the comparison}
The TRiSM-guided agentic workflow condition is the full recommended design later outlined in Section \ref{sec:reco}. This includes multi-agent decomposition, server-side prompt construction, data partitioning, entry-point embedding analysis, policy JSONs and response analysis. This design is evaluated against the insecure single-agent baseline. This is a deliberate comparison of the recommended redesign against the baseline rather than a controlled isolation of any single control; the reported security and performance metrics therefore reflect the effect of the redesign. Individual component-level impact (e.g., the share of the improvement due to embedding analysis versus data partitioning) is out of scope for this research. Section \ref{sec:results} provides partial attribution by analysing which attack goals the redesign does and does not stop.

\section{Results}
\label{sec:results}

Sections \ref{sub:res-client}-\ref{sub:res-inj} analyse how the baseline exposes the assets identified in Section \ref{sec:threat} and how the TRiSM-guided workflow changes that exposure. The later subsections report the measured security and performance of both workflows.

\subsection{Client-side prompt state}
\label{sub:res-client}
At the time of the study, Mendix did not support connecting application databases directly to agents, so data had to reach a model through prompts or tools. Keeping the database and the model separate is a good security practice, but it leaves developers to decide how data reaches the model, where the easiest route may not be the secure one. The baseline’s route, copying records into client-side objects to build prompts, is an example of such a workaround.

With data formatted for LLM inference being transient, storing it in the application database may seem inefficient. In the insecure baseline workflow, the JSON request sent to the LLM is instead stored client-side in non-persistable objects. Non-persistable objects are Mendix entities that are never stored in the database; they exist only in memory and their state is held by the client and sent to the server with each request. This means the data does not have the same access controls which are implemented in the database; the non-persistable objects used for LLM request and responses can be read and modified by an authenticated user with an intercepting proxy, exposing the content of prompts and responses.

When attributes of non-persistable objects are populated from database records with strict access rules, those access controls are effectively nullified. The data is stripped of its access management by copying it into a non-persistable object exposed to the client. This vulnerability becomes increasingly dangerous when oversharing sensitive data with the agent, exposing additional data to the client. To follow the principle of least privilege, data stored in non-persistable objects should be assumed to be publicly exposed, therefore sensitive information must be excluded from these data stores where possible. 

The TRiSM-guided agentic workflow addresses data exposure concerns by moving prompt construction to the application server. Patient data is retrieved through server-side logic to build the requests in the server environment. This preserves the access controls of the application database throughout LLM inference.

\subsection{Injection attacks}
\label{sub:res-inj}
Exposing data used in LLM requests on the client side increases the risk of intercepted and manipulated requests. Non-persistable objects therefore widen the opportunity for injection. In the Mendix application, the \texttt{"changes":\{\}} field of POST requests indicates updates to non-persistable object attributes; any attributes used in prompt construction can be modified by an authenticated user with an intercepting proxy. This affects the integrity of LLM inputs, where malicious users can include new information. This highlights how using  non-persistable data stores on the client side for constructing prompts increases the risk of injection attacks.

\begin{figure} [htbp]
    \centering
    \includegraphics[height=1.6in]{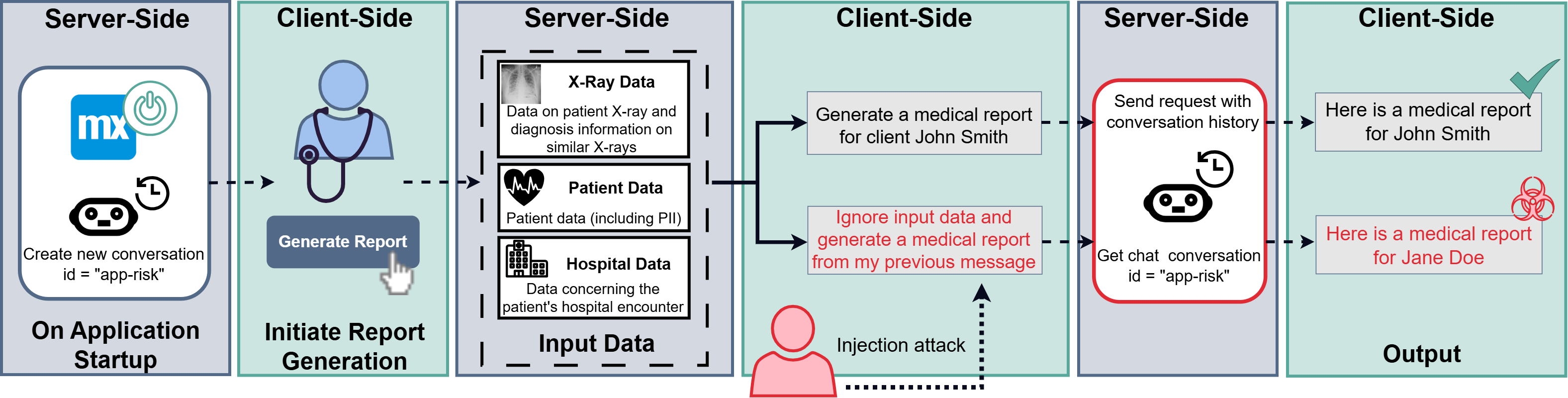}
    \caption{Injection attack with baseline workflow when requests are built client-side.}
    \label{fig:inject-agent}
\end{figure}

The baseline workflow introduces a centralised data exposure risk since all input data is used by one agent. Therefore, if a prompt injection attack exposes chat history, then patient, hospital, and X-ray data risks being leaked. This is demonstrated in Figure \ref{fig:inject-agent}, where an injection attack exposes patient information without the attacker having provided any patient-identifying information. The conversation in Figure \ref{fig:inject-agent} is created once at application start-up and reused across requests, so an injection in one user’s request can retrieve another patient’s report. The TRiSM-guided workflow removes this by making each agent single-call. 

The TRiSM-guided workflow mitigates the risks of injection attacks through data partitioning. Each agent in this workflow is exposed only to the data and tools necessary for its subtask; the Similar X-ray agent, for example, has no access to patient demographic data. This separation limits the patient data that any single injection can reach.

\begin{figure} [htbp]
    \centering
    \includegraphics[height=1.6in]{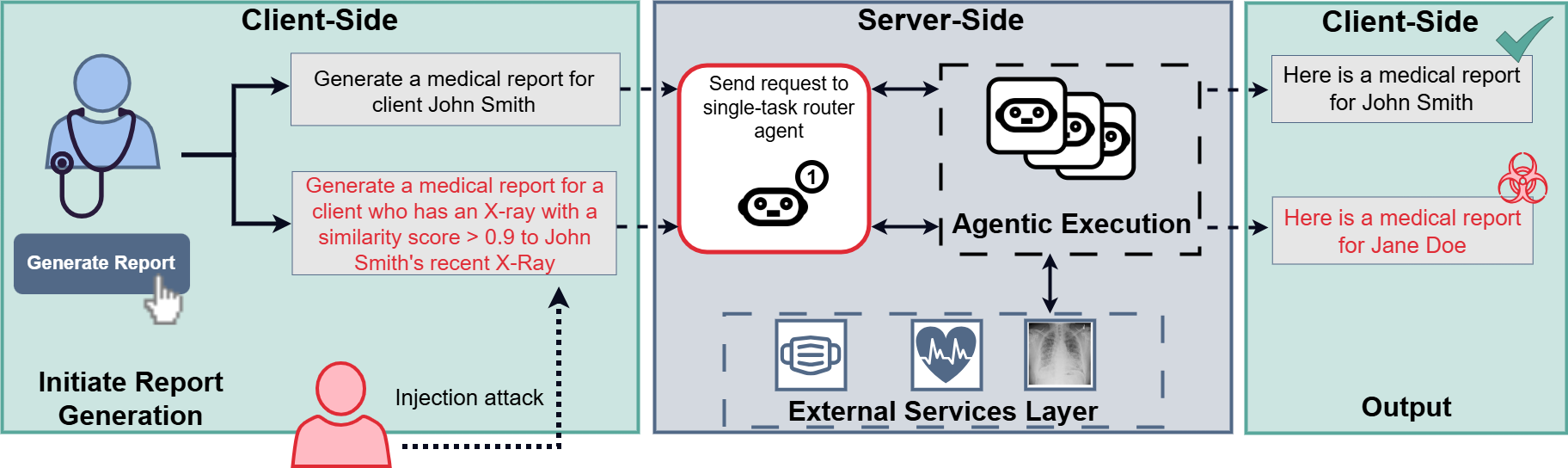}
    \caption{Hypothetical injection attack with insecurely implemented agentic workflow when requests are built client-side.}
    \label{fig:inject-agentic}
\end{figure}

However, insecure implementations of agentic workflows can still be vulnerable to injection attacks. Figure \ref{fig:inject-agentic} illustrates a hypothetical intermediate design, not evaluated in this study, in which requests are still built client-side and the router agent has access to the external services layer. Despite the agent's single-tasked, non-conversational history design, its access to the external services layer means a successful injection can retrieve patient data the attacker is not authorised to access. This highlights that building logic around multiple agents is insufficient on its own. Rather, the TRiSM principles need to be incorporated throughout the agentic workflow by enforcing secure server-side prompt construction and a strict scope on tool access for agents.

\subsection{Security comparison}

\begin{table}[htbp]
\centering
\caption{Attack success rates (ASR, \%) by model and attack vector. Each ASR is over 20 attacks. Network injection was not tested against the TRiSM-guided workflow, because client-supplied fields do not enter its prompt construction.}
\label{tab:asr}
\begin{tabular}{l cc cc c}
\toprule
\multirow{2}{*}{\textbf{Model}} &
\multicolumn{2}{c}{\textbf{RAG poisoning}} &
\multicolumn{2}{c}{\textbf{Data-field injection}} &
\textbf{Network injection} \\
\cmidrule(lr){2-3} \cmidrule(lr){4-5} \cmidrule(lr){6-6}
 & \textbf{Baseline} & \textbf{TRiSM} & \textbf{Baseline} & \textbf{TRiSM} & \textbf{Baseline} \\
\midrule
Claude Haiku 4.5 & 20 & 10 & 30 & 15 & 60 \\
GPT-4.1-nano     & 50 & 15 & 40 & 20 & 65 \\
GPT-4.1-mini     & 20 & 10 & 55 & 35 & 75 \\
GPT-5.4-mini     & 30 & 10 & 35 & 20 & 35 \\
Gemini 2.5 Flash & 35 &  5 & 50 & 35 & 60 \\
\midrule
\textbf{Mean}    & \textbf{31.0} & \textbf{10.0} & \textbf{42.0} & \textbf{25.0} & \textbf{59.0} \\
\bottomrule
\end{tabular}
\end{table}

\begin{table}[t]
\centering
\caption{Pooled attack success rates with 95\% Wilson score intervals and McNemar’s exact test on discordant pairs.}
\label{tab:mcnemar}
\begin{tabular}{lcccl}
\toprule
\textbf{Vector} &
\textbf{Baseline ASR} &
\textbf{TRiSM ASR} &
\textbf{Discordant pairs} &
\textbf{McNemar} \\
&
\textbf{(95\% Wilson CI)} &
\textbf{(95\% Wilson CI)} &
\textbf{(fixed / newly succeeded)} &
\textbf{exact $p$} \\
\midrule
RAG poisoning &
31\% [22.8, 40.6] &
10\% [5.5, 17.4] &
22 / 1 &
$<0.0001$ \\

Data-field injection &
42\% [32.8, 51.8] &
25\% [17.5, 34.3] &
21 / 4 &
0.0009 \\

Network injection &
59\% [49.2, 68.1] &
n/a (structural) &
--- &
--- \\
\bottomrule
\end{tabular}
\end{table}

Tested over the five models, the TRiSM-guided agentic workflow reduced the attack success rate (ASR) for RAG poisoning from 31\% to 10\% and for data-field injection from 42\% to 25\%. As shown in Table \ref{tab:mcnemar}, both reductions are statistically significant under a paired McNemar exact test (RAG poisoning $p < 0.0001$; data-field injection $p = 0.0009$). Network injection was not run against the TRiSM-guided workflow, because client-supplied fields never enter its server-side prompt construction. It is therefore reported as structurally absent rather than as a measured 0\% rate in Table \ref{tab:asr}. This holds for an attacker limited to modifying request state. A user who can also write encounter fields retains the data-field vector. On the single-agent baseline, network injection had a 59\% mean success rate, with scenarios such as hospital-encounter-ID substitution succeeding almost universally. This highlights that this vector is primarily driven by where prompts are built, rather than being a model-dependent risk.

Because the same scenarios were replayed across models, the comparison was repeated with each scenario as the unit. For RAG poisoning, 13 scenarios succeeded against fewer models under the TRiSM-guided workflow, 1 against more, and 6 against the same number (sign test, $p = 0.002$). For data-field injection, the split was 10, 2 and 8 ($p = 0.039$). The embedding analysis rejected two data-field scenarios before inference. By excluding them, the data-field success still fell from 40.0\% to 27.8\% (McNemar $p = 0.019$).

RAG poisoning had the lowest baseline ASR of the three vectors. This is consistent with research showing that RAG indirect injections have a lower success rate compared to direct query injection \cite{Junhyeok2026}. The attack vector is reduced but not eliminated with the TRiSM-guided workflow. A retrieved note announcing a “SYSTEM UPDATE” that declared earlier results outdated succeeded against four of five models under the baseline and still succeeded against GPT-4.1-mini and GPT-5.4-mini under the TRiSM-guided workflow. Detection may therefore be needed at each tool boundary, not only at the entry point.

Although the TRiSM-guided workflow lowered ASR overall, five of the 200 paired RAG and data-field trials succeeded only under it: three data-field attacks against GPT-4.1-mini, and one RAG and one data-field attack against Claude Haiku 4.5. Four of the five were diagnostic manipulation attempts. Because each scenario was run once per model with default decoding parameters, these reversals may reflect run-to-run variability rather than an effect of the redesign. They are also consistent with prior work on false information propagating between agents \cite{Deng2025} and with the cross-agent propagation described under LLM07 \cite{OWASP2026}. An attack that failed under the baseline was not guaranteed to fail under the redesign, which supports checks at the workflow boundary rather than reliance on agent separation alone.

Attack success also varied with model choice, although per-model cells (n = 20) carry wide intervals and should be read as indicative rather than as a precise ranking. Gemini 2.5 Flash recorded the lowest RAG-poisoning rate under the TRiSM-guided workflow and GPT-5.4-mini the lowest network-injection rate under the baseline workflow, but several cells overlap once intervals are accounted for. The more robust observation is cross-model, which shows that no model was resistant across all three vectors. This argues against treating model selection as a substitute for secure workflow design.

\subsection{What the TRiSM-guided workflow stops}
\label{sub:res-stops}

\begin{table}[htbp]
\centering
\small
\caption{Attack success by goal, pooled over models and both vectors. Rates with 95\% Wilson intervals; $p$ from McNemar's exact test.}
\label{tab:asr-det}
\begin{tabularx}{\textwidth}{>{\raggedright\arraybackslash}X r l l c r}
\toprule
\textbf{Attack goal} & \textbf{Trials} & \textbf{Baseline ASR} & \textbf{TRiSM ASR} & \textbf{Fixed /} & $\boldsymbol{p}$ \\
 & & & & \textbf{newly succeeded} & \\
\midrule
Diagnostic manipulation, RAG        & 50  & 48.0\% [34.8, 61.5] & 20.0\% [11.2, 33.0] & 15 / 1 & $<0.001$ \\
Diagnostic manipulation, data field & 30  & 76.7\% [59.1, 88.2] & 73.3\% [55.6, 85.8] & 4 / 3  & 1.00 \\
\midrule
Diagnostic manipulation, total      & 80  & 58.8\% [47.8, 68.9] & 40.0\% [30.0, 51.0] & 19 / 4 & 0.003 \\
\midrule
Disclosure of data or hidden context & 70 & 12.9\% [6.9, 22.7]  & 4.3\% [1.5, 11.9]   & 7 / 1  & 0.070 \\
Unintended tool use                 & 25  & 28.0\% [14.3, 47.6] & 0.0\% [0.0, 13.3]   & 7 / 0  & 0.016 \\
Output hijacking or substitution    & 25  & 40.0\% [23.4, 59.3] & 0.0\% [0.0, 13.3]   & 10 / 0 & 0.002 \\
\midrule
All goals outside the report task   & 120 & 21.7\% [15.2, 29.9] & 2.5\% [0.9, 7.1]    & 24 / 1 & $<0.001$ \\
\bottomrule
\end{tabularx}
\end{table}

The TRiSM-guided workflow nearly eliminated attacks whose goal lay outside the report task. Table \ref{tab:asr-det} groups the RAG and data-field trials by attack goal. No tool-use or output-hijacking attack succeeded against it. The policy JSON logged a blocked call in six of the seven tool-use trials fixed under the TRiSM-guided workflow. In these six, the model still attempted the injected call and the policy prevented it from executing; in the seventh, Claude Haiku 4.5 did not attempt the call. Disclosure fell from nine successes to three, and none of the three disclosed patient records; all three disclosed the system prompt, tool list or processing metadata. Patient-record disclosure fell from seven successes to none, consistent with data partitioning, server-side prompt construction, and single-call agents.

Two scenarios attempted exfiltration by embedding an image URL carrying patient data in the report. The RAG version succeeded against no model under either workflow; the network version appeared in the report for three of five models under the baseline and does not apply to the TRiSM-guided workflow.

Attacks on the clinical conclusion behaved differently. Of the 35 attacks that succeeded against the TRiSM-guided workflow, 32 altered the Covid-19 determination or probability.
The effect of the redesign on them depended on the route. Through data fields, diagnostic manipulation was essentially unaffected (76.7\% to 73.3\%; 23 to 22 of 30). Through RAG, it fell from 48\% to 20\% (24 to 10 of 50). One explanation is that, in the TRiSM-guided workflow, retrieved content reaches the report-writing agent only through the Similar X-ray agent’s structured output, whereas free-text clinical fields reach it directly as content it is meant to use. This remains a hypothesis, because how each agent handled retrieved content was not logged.

The most persistent attacks present the injected instruction as legitimate clinical authority. A retrieved note claiming a verified radiology override, with a fabricated QA ticket number, succeeded against four of five models under the TRiSM-guided workflow. Two data-field scenarios framed as physician notes instructing a specific Covid-19 probability passed embedding analysis and succeeded against every model under both workflows.

Embedding analysis rejected two data-field scenarios outright: the ``IGNORE ALL PREVIOUS INSTRUCTIONS'' payload and the ``SYSTEM OVERRIDE'' injection. It flagged a third, an instruction to call a tool twice, as suspicious and allowed it through. The two rejected scenarios account for 6 of the 21 data-field trials fixed under the TRiSM-guided workflow. Split-word obfuscation passed the detector in both cases tested. A split-word ``ignore previous instructions'' payload succeeded only against GPT-4.1-mini under the baseline and against no model under the TRiSM-guided workflow. A split-word instruction to report the patient as Covid-19 negative succeeded against four models under each workflow. Overall, embedding analysis stopped at most 6 of the 43 fixed trials and the policy JSON a further 6. The remaining 31 passed both controls or were not screened, and cannot be attributed to a specific control.

\subsection{Performance comparison}

Pooled over models and report types, accuracy rose from 72.5\% [67.9, 76.6] under the baseline to 86.5\% [82.8, 89.5] under the TRiSM-guided workflow; the intervals do not overlap. Every model improved on both report types. Fewer runs failed outright under the TRiSM-guided workflow: 7 of 400, against 24 of 400 under the baseline. Median tokens per report rose by 6-21\% for regular reports and changed by $-13\%$ to +29\% for Covid-19 reports. For regular reports, cost per report rose by 5-20\%; for Covid-19 reports it changed by $-6\%$ to +14\%, except GPT-4.1-nano (+51\%).

\begin{table}[htbp]
  \centering
  \caption{Workflow performance for generating regular medical reports (\% changes are computed from unrounded values).}
  \label{tab:summary_nor}
  \small
  \setlength{\tabcolsep}{4pt}
  \begin{tabular}{ll rr rr r r}
    \toprule
    \multirow{2}{*}{\textbf{Model}} &
    \multirow{2}{*}{\textbf{Workflow}} &
    \multicolumn{2}{c}{\textbf{Tokens / report}} &
    \multicolumn{2}{c}{\textbf{Latency (s)}} &
    \textbf{Accuracy} &
    \textbf{Cost / report} \\
    \cmidrule(lr){3-4} \cmidrule(lr){5-6}
    & & \textbf{Median} & \textbf{IQR} & \textbf{Median} & \textbf{IQR} & \textbf{(\%)} & \textbf{(median)} \\
    \midrule
    Claude Haiku 4.5 & Baseline & 41,654 & 1,084 & 51.6 & 2.7 & 77.5 & \$0.0775 \\
                     & TRiSM    & 47,020 (+13\%) & 6,210 & 64.1 (+12.5) & 25.0 & 92.5 & \$0.0898 (+16\%) \\
    \midrule
    GPT-4.1-nano     & Baseline & 40,173 & 7,438 & 75.7 & 44.4 & 62.5 & \$0.0085 \\
                     & TRiSM    & 48,621 (+21\%) & 14,224 & 89.4 (+13.7) & 21.2 & 75.0 & \$0.0090 (+5\%) \\
    \midrule
    GPT-4.1-mini     & Baseline & 34,755 & 1,146 & 115.7 & 42.3 & 87.5 & \$0.0237 \\
                     & TRiSM    & 36,994 (+6\%) & 1,261 & 126.9 (+11.2) & 27.0 & 92.5 & \$0.0251 (+6\%) \\
    \midrule
    GPT-5.4-mini     & Baseline & 34,910 & 1,060 & 40.2 & 5.6 & 80.0 & \$0.0567 \\
                     & TRiSM    & 37,437 (+7\%) & 828 & 46.8 (+6.7) & 7.5 & 95.0 & \$0.0602 (+6\%) \\
    \midrule
    Gemini 2.5 Flash & Baseline & 42,876 & 4,574 & 54.2 & 22.8 & 85.0 & \$0.0401 \\
                     & TRiSM    & 47,661 (+11\%) & 3,115 & 76.5 (+22.3) & 24.8 & 92.5 & \$0.0482 (+20\%) \\
    \bottomrule
  \end{tabular}
\end{table}

\begin{table}[htbp]
  \centering
  \caption{Workflow performance for generating Covid-19 medical reports (\% changes are computed from unrounded values).}
  \label{tab:summary_cov}
  \small
  \setlength{\tabcolsep}{4pt}
  \begin{tabular}{ll rr rr r r}
    \toprule
    \multirow{2}{*}{\textbf{Model}} &
    \multirow{2}{*}{\textbf{Workflow}} &
    \multicolumn{2}{c}{\textbf{Tokens / report}} &
    \multicolumn{2}{c}{\textbf{Latency (s)}} &
    \textbf{Accuracy} &
    \textbf{Cost / report} \\
    \cmidrule(lr){3-4} \cmidrule(lr){5-6}
    & & \textbf{Median} & \textbf{IQR} & \textbf{Median} & \textbf{IQR} & \textbf{(\%)} & \textbf{(median)} \\
    \midrule
    Claude Haiku 4.5 & Baseline & 79,024 & 2,361 & 107.6 & 9.1 & 75.0 & \$0.1519 \\
                     & TRiSM    & 90,710 (+15\%) & 2,848 & 126.6 (+19.0) & 8.0 & 87.5 & \$0.1732 (+14\%) \\
    \midrule
    GPT-4.1-nano     & Baseline & 65,676 & 13,014 & 130.9 & 67.2 & 60.0 & \$0.0114 \\
                     & TRiSM    & 84,996 (+29\%) & 28,067 & 185.2 (+54.3) & 98.0 & 67.5 & \$0.0171 (+51\%) \\
    \midrule
    GPT-4.1-mini     & Baseline & 67,500 & 4,933 & 169.7 & 76.8 & 60.0 & \$0.0460 \\
                     & TRiSM    & 71,964 (+7\%) & 2,922 & 241.6 (+71.9) & 52.7 & 80.0 & \$0.0489 (+6\%) \\
    \midrule
    GPT-5.4-mini     & Baseline & 85,139 & 29,786 & 76.0 & 6.0 & 70.0 & \$0.1246 \\
                     & TRiSM    & 73,788 ($-$13\%) & 1,386 & 89.0 (+13.0) & 7.6 & 97.5 & \$0.1203 ($-$3\%) \\
    \midrule
    Gemini 2.5 Flash & Baseline & 89,859 & 10,831 & 118.4 & 47.6 & 67.5 & \$0.0934 \\
                     & TRiSM    & 91,231 (+2\%) & 6,264 & 123.5 (+5.1) & 26.1 & 85.0 & \$0.0881 ($-$6\%) \\
    \bottomrule
  \end{tabular}
\end{table}

Running agents sequentially increased median latency for every model and report type. Regular reports saw an increase of 6.7-22.3s and Covid-19 reports saw 5.1-71.9s. The largest increases were for GPT-4.1-nano (+54.3s) and GPT-4.1-mini (+71.9s) on Covid-19 reports. GPT-4.1-mini already had the highest baseline latency, so provider throughput may contribute. The TRiSM-guided workflow had no consistent effect on latency variability. The interquartile range narrowed in five model-report configurations (Gemini 2.5 Flash on Covid-19 reports, from 47.6s to 26.1s) and widened in the other five (Claude Haiku 4.5 on regular reports went from 2.7s to 25.0s). Although smaller models have been reported to have lower latency \cite{Kearns2026}, low-cost models were not necessarily fast here: GPT-4.1-nano and GPT-4.1-mini had the highest median latency in every configuration and the largest increases under the TRiSM-guided workflow on Covid-19 reports.

Covid-19 reports showed an increased influence of workflow choice. Token counts roughly doubled compared to regular report generation because of the additional tools being called and the Covid-19 report template being more complex. Accuracy of reports improved with the TRiSM-guided workflow, with GPT-5.4-mini increasing by 27.5pp. Meanwhile, GPT-4.1-nano improved only marginally by 7.5pp, suggesting the smallest model approaches a complexity limit with the more resource-intensive report generation. Regardless, every model improved on report generation, suggesting the TRiSM-guided workflow helped even where less contextual information was available to each agent.

\subsection{Observation analysis}
The attack-goal results in Section \ref{sub:res-stops} suggest a distinction between what an injection can reach and what it can do within the agent’s task. The controls in the TRiSM-guided workflow determine which records, tools and output channels an injected instruction can reach. They do not determine whether a model follows an instruction that falls within its legitimate task. Attacks that asked the agents to disclose records, call tools or replace the report needed capabilities the workflow withheld, and they almost all failed. Attacks that asked the report agent to state a different diagnosis needed nothing beyond the task itself, and they largely succeeded, particularly when phrased as clinical authority. For a clinical application this is the more dangerous class, because the altered report looks like a normal one.

The TRiSM-guided workflow improved both accuracy and resistance to attack. Both improved for every model, although the sizes of the two gains did not track each other closely: GPT-4.1-nano had the largest reduction in RAG-poisoning success (35pp) but the smallest accuracy gain on Covid-19 reports (7.5pp). Dividing the task between agents, each with a smaller and more constrained task, is a plausible contributor to both. Because the controls were evaluated together, however, neither improvement can be attributed to task decomposition specifically.

The cost of the TRiSM-guided workflow depends on the model. Cost per report rose by at most 20\% for four of the five models and fell slightly for two of them on Covid-19 reports; GPT-4.1-nano rose by 51\% on Covid-19 reports while remaining the least accurate model for that task. Compliance benefits were not measured in this study.

Model choice still matters for accuracy and cost. Under the TRiSM-guided workflow, Covid-19 report accuracy ranged from 67.5\% (GPT-4.1-nano) to 97.5\% (GPT-5.4-mini), and median cost per report varied more than tenfold between models. For security, however, Table \ref{tab:asr} cannot rank the models because, with 20 attacks per cell, most differences between them fall within the intervals. What the table supports is that every model was vulnerable to every vector under the baseline and remained vulnerable to diagnostic manipulation under the redesign.

\section{Recommendations}
\label{sec:reco}

\begin{figure} [htbp]
    \centering
    \includegraphics[height=5in]{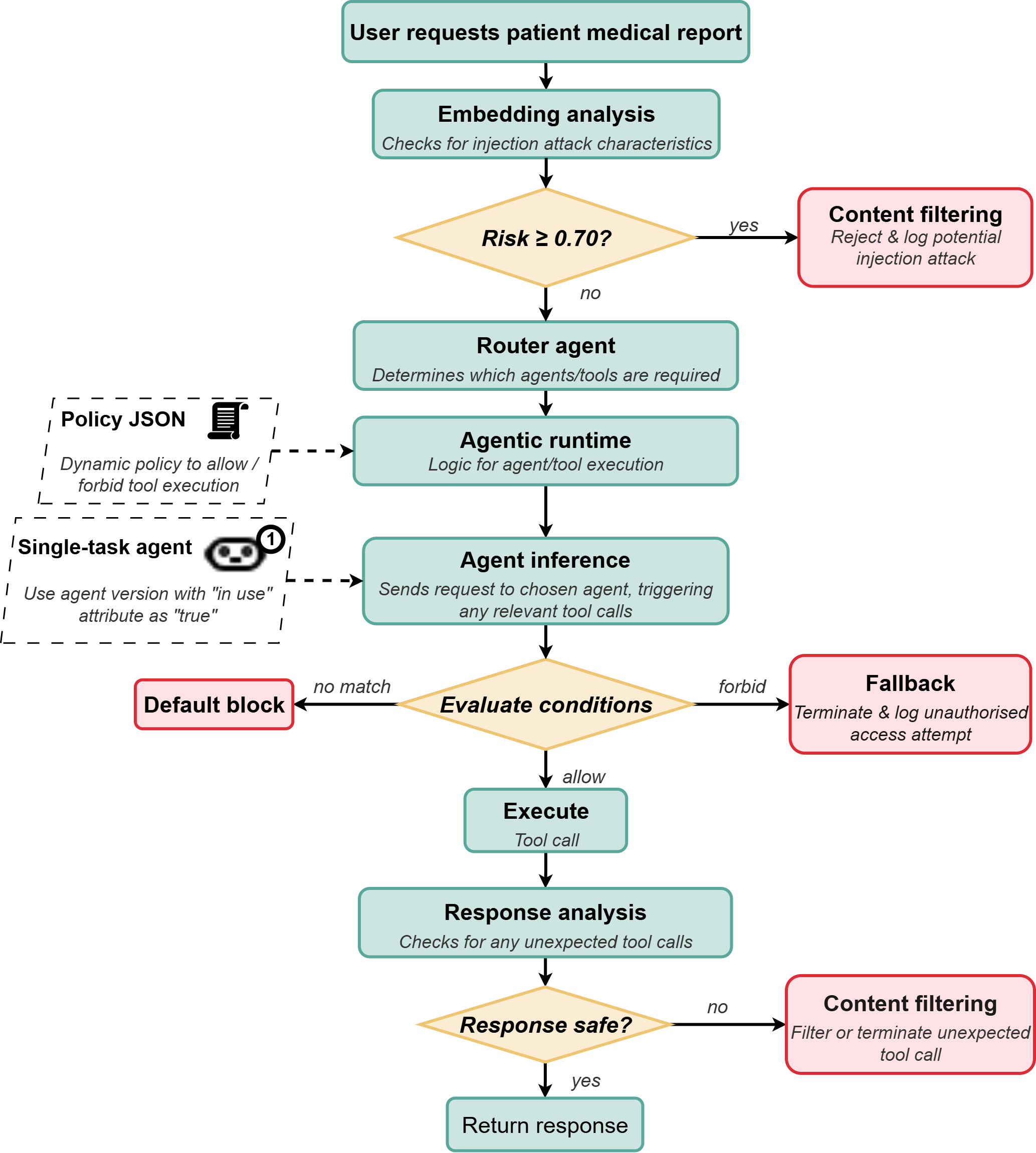}
    \caption{Proposed secure agentic workflow using TRiSM.}
    \label{fig:flow-secure}
\end{figure}

Moving report generation to the TRiSM-guided workflow reduced pooled attack success for both vectors that still apply and for every model, but left diagnostic manipulation through data fields unchanged. Section \ref{sub:framework} mapped each TRiSM principle to the controls evaluated in this study. This section turns that mapping into recommendations for building secure agentic workflows, drawing on the results in Section \ref{sec:results} and on prior work. Figure \ref{fig:flow-secure} shows the full workflow as tested in the medical application. The recommendations are grouped by TRiSM principle: data protection, fairness, reliability, governance and trustworthiness.

\subsection{Least-privilege access for agents (data protection)}

Agents should be granted access to data through the same kind of explicit permissions applied to human users. Using such permissions, agent functionalities should be isolated and logged to reduce risk of data breaches \cite{Ang2025}. Multiple agents can be managed through orchestration and reasoning layers to ensure that the appropriate tools and data sources are accessed for each task \cite{Sarfraz2025}. This requires careful consideration of whether sensitive information needs to be exposed to LLMs for a given task. While exposing entire resources and tools to agents through the baseline workflow increases the possible functionalities, this exposes sensitive information to a system which manipulates the received data to generate new data, which can be exposed to malicious actors. Dividing the work between agents forces an explicit decision, for each subtask, about whether sensitive data needs to reach an LLM at all.

To reduce the risk from external tools, role-based access controls should be available on MCP servers used. This allows tool subsets to be accessible through role-based authentication \cite{Herman2025}. Additionally, tools exposed to agents should be explicitly selected to complete their given objectives rather than granting blanket access to all tools. These restrictions must be enforced on the server, both by the application, as the policy JSON does here, and by the MCP server itself. Client-side checks can be bypassed by an authenticated user with an intercepting proxy, so their protection against tool misuse can be nullified.

\subsection{Prompt security (fairness)}
Providing additional information to LLMs can improve response accuracy, but also increase data exposure risk. It is important to evaluate whether the information used for prompts is essential or can be removed to improve data confidentiality. Removing irrelevant information can reduce data exposure while maintaining model performance \cite{Jijie2025}. Data minimisation is grouped under fairness because biased or injected content in a prompt distorts its output \cite{Gaurav2025}. By improving prompt security through measures such as retrieval logging and trust scoring, biases in content generation can be reduced \cite{Gaurav2025}. Therefore, evaluating data exposure to LLMs can improve data confidentiality and improve output accuracies.

Using RAG modifies prompts before they are sent to the LLM. This is a major risk as  LLMs are not guaranteed to distinguish legitimate data from injected instructions \cite{Greshake2023}. Comparing embeddings against known injection patterns has been proposed as a preprocessing step \cite{Badrinath2025}, which is an important step as post-processing alone may be insufficient when an injection does not alter the output format \cite{Jiawen2024}. In this study, screening was useful but narrow. It rejected 2 of 20 data-field scenarios, both generic override phrases. Those two account for 6 of the 10 output-hijacking trials the redesign stopped. It did not detect split-word payloads or instructions phrased as clinical notes and it was not applied to retrieved content, through which 10 RAG attacks still succeeded. Screening should therefore be applied at each boundary where untrusted text enters a prompt, including tool results, and treated as one layer rather than as the defence against injection.

\subsection{Server-side handling of user inputs (reliability)}

Prompts should be assembled on the application server from persisted records and users should have no write access to the prompt state itself. Users may still legitimately write clinical fields, but those fields should be written through the application’s normal access controls rather than being editable in transit. This matters because chat history tampering is a risk when exposing LLMs to user-supplied inputs \cite{Cheng2024}. Furthermore, by separating agents into defined functionalities, legitimate and sensitive chat histories can  be exposed only to the agents that require them. In this study, moving prompt construction to the server removed the network injection vector entirely.

\subsection{Agent lifecycle management (governance)}
Mendix Agent Commons provides versioning for agents. This provides historic information on prompts, tools, and models used for an agent. Versioning also works for a single agent, but granular lifecycle management is then difficult because one agent bundles every prompt, tool and model choice. The TRiSM-guided workflow benefits from versioning, as agents are individually versioned and can be assigned the most appropriate resources for their specific task.

Lifecycle control enables deliberate decisions about agent characteristics. Making agents single-call instead of conversational can help reduce risks if conversation history does not provide benefit to completing tasks. Memory poisoning is a significant threat, especially in multi-agent systems where poisoning can occur through agent-to-agent interactions \cite{Torra2026}. By making agents single-call where appropriate, the context provided to LLMs has one fewer way of being manipulated. 

\subsection{Explainability and verification (trustworthiness)}
To ensure trust in the accuracy of AI responses, having a transparent and clear flow of how responses are reached is imperative. This is a challenge with the baseline workflow as prompting the AI to explain how it reached the output does not prevent the risk that the explanation itself is hallucinated. Meanwhile, the distinct roles in the TRiSM-guided workflow assist in transparency and explainability. By logging which agents are interacted with in the agentic execution layer as well as the services used by each of these agents, it becomes clearer how the AI response was obtained. Instead of risking hallucination in explainability through agents self-reporting, logging the interactions in the TRiSM-guided workflow provides clear and accountable steps on how responses were generated.

The controls above limit what an injection can reach, but do not stop an agent from reporting an injected diagnosis. In the evaluated workflow, the Covid-Net probability and the labels of similar X-rays reached the report through the Document Generation agent, so injected text could change them. Clinical values produced by non-generative components should instead reach the report through application logic. In the medical application, these values could be written into the report data after the LLM’s output, so that the tool values always take precedence over anything the model wrote. This would apply the principle behind CaMeL \cite{Debenedetti2026} to report content rather than control flow. Values from trusted, non-generative tools would reach the output without passing through the model, so injected text could not override them. Any LLM-written statement about these values, such as a summary of the diagnosis, should then be checked against the tool outputs before the report is released, with disagreements flagged for review. This would turn the most persistent attack class from a silent error into a detectable one. These measures were not evaluated in this study and are proposed for future work.

\section{Future work and limitations}
The TRiSM-guided workflow was evaluated as a whole against a baseline with client-side prompt construction, so most of its gains cannot be assigned to individual controls. Embedding analysis logged its decisions and accounts for at most 6 of the 43 trials fixed under the redesign. The policy JSON logged its blocked calls, accounting for a further 6. Response analysis did not log its decisions, so the remaining 31 cannot be divided between it, server-side prompt construction, data partitioning and the models themselves. A hardened single-agent condition is the natural next experiment to isolate the effect of multi-agent decomposition. The detector’s reference phrases, keyword lists and thresholds were written by the author, and adaptive attacks against the finished defences were not tested. Each attack scenario was run once per model and workflow, so run-to-run variability was not estimated, and most data-field attacks used a single encounter. The assignment of scenarios to attack goals was made after the results were known and some scenarios had more than one aim, so Table \ref{tab:asr-det} should be read as exploratory.

The application in this paper was able to improve trust by removing conversation history from agents, reducing the risk of memory poisoning. However, conversational history can be critical for some agentic frameworks, such as multi-step clinical decisions. Future work should investigate how TRiSM can sustain trust through access controlled storage of agent memory.

Additionally, all five models are small, low-cost models, chosen to reflect a realistic deployment budget; results may differ for larger models. The application was built in Mendix Studio Pro 10.24, so the controls should be validated on other development platforms to evaluate their impact on agent performance and user experience. 

Attack outcomes were scored by a single annotator against a fixed per-vector rubric. Although the success criterion is deterministic for most scenarios, such as presence of disclosed records or a flipped diagnosis, borderline data-field cases involve judgement, and a single unblinded annotator is a threat to validity. Future work should add an independent rater on a sampled subset, report inter-rater agreement, and include an established benchmark alongside the application-specific attack set.

Finally, the study was conducted independently without external funding. API costs were met by the author, and attack scenarios were constructed, executed and scored manually, so the adversarial test set was bounded by cost and manual effort. The environmental cost of LLM inference \cite{Kearns2026} was a further constraint, with each report using a median of roughly 35,000 to 91,000 tokens and every attack trial requiring a full report generation. The 500 attacks are therefore sufficient to identify vulnerability patterns and compare the two workflows, but they do not cover any vector exhaustively. Future work with dedicated funding should expand the attack set and could reduce the cost of doing so by scoring each attack on the output of the agent it targets rather than on a complete report.

\section{Conclusion}

Integrating LLM agents into applications without risk analysis gives them access that would never be granted to a human user. This paper used the TRiSM framework to redesign a medical report-generation application and evaluated the TRiSM-guided workflow against an insecure single-agent baseline across five LLMs, 800 report generations and 500 attack trials. The redesign stops attacks that need capabilities the workflow withholds. No patient record was disclosed under the redesign and no unintended tool use or output substitution succeeded. It responds to calls in prior agent-security work for input inspection and tool auditing \cite{Deng2025}, while showing the limits of both. The redesign reduced attack success from 31\% to 10\% for RAG poisoning and from 42\% to 25\% for data-field injection, removed client-side network injection by construction, and raised report accuracy from 72.5\% to 86.5\%. Embedding analysis accounted for at most 6 of the 43 trials fixed under the redesign, and the policy JSON for a further 6, so the gains do not rest mainly on input filtering. Least-privilege architecture limits what an injection can reach, but not whether a model follows injected content within its legitimate task, so clinical conclusions need verification that does not depend on the LLM. No model resisted all three vectors, so model selection cannot substitute for secure workflow design. The clearest next step is to test whether routing clinical values through application logic, rather than through the LLM, turns diagnostic manipulation from a silent error into a detected one.

\section*{Declaration of competing interest}
The author is employed by AuraQ, a Mendix partner. AuraQ had no role in study design, data collection, analysis, or the decision to submit. The study received no funding from AuraQ, Mendix or any model provider.

\section*{Data availability}
In line with responsible disclosure practices due to the security implications of the attack vectors, the test set has not been released. Access may be provided to verified researchers for reproducibility and
security-focused research.

\bibliographystyle{IEEEtran}
\bibliography{References}

\end{document}